\documentclass[a4paper]{article}
\usepackage{array,amsfonts,graphicx, xcolor, graphics}
\usepackage{INTERSPEECH2020}
\usepackage{libertine}
\usepackage{amsmath,booktabs,nccmath}
\usepackage{multirow,array,makecell}
\usepackage[font=small]{caption}
\usepackage[nopar]{lipsum}
\usepackage{diagbox}
\usepackage{subcaption}

\usepackage[activate]{microtype}
\pdfoutput=1

\title{Deep Architecture Enhancing Robustness to Noise, Adversarial Attacks, and Cross-corpus Setting for Speech Emotion Recognition}
\name{Siddique Latif$^{1,2}$, Rajib Rana$^1$, Sara Khalifa$^{2,3}$,  Raja Jurdak$^4$, Bj\"{o}rn W.\ Schuller$^{5,6}$}

\address{
  $^1$University of Southern Queensland, Australia\\
  $^2$Distributed Sensing Systems Group, Data61, CSIRO Australia\\
  $^3$University of New South Wales, Australia\\
  $^4$Queensland University of Technology, Australia\\
  $^5$GLAM -- Group on Language, Audio, \& Music, Imperial College London, UK\\
  $^6$Chair of Embedded Intelligence for Health Care and Wellbeing, University of Augsburg, Germany}
\email{siddique.latif@usq.edu.au}

\begin{document}

\maketitle
\begin{abstract}

Speech emotion recognition systems (SER) can achieve high accuracy when the training and test data are identically distributed, but this assumption is frequently violated in practice and the performance of SER systems plummet against unforeseen data shifts. The design of robust models for accurate SER  is challenging, which limits its use in practical applications. In this paper we propose a deeper neural network architecture wherein we fuse DenseNet, LSTM and Highway Network to learn powerful discriminative features which are robust to noise. We also propose data augmentation with our network architecture to further improve the robustness.  We comprehensively evaluate the architecture coupled with data augmentation against (1) noise, (2) adversarial attacks and (3) cross-corpus settings. Our evaluations on the widely used IEMOCAP and MSP-IMPROV datasets show promising results when compared with existing studies and state-of-the-art models.

% which helps to learn powerful discriminitive features improving the robustness of SER. To further improve the robustness we   limite to use automatic speech emotion recognition in practical applications accuracy needs to be significantly increased. the accuracies reported in the literature i. Based on the previous success of models using convolutional neural networks (CNNs) and subsequent recurrent neural networks (RNNs) and general deep neural networks  (DNNs) in the same network, we propose to use a 
% densely connected convolutional network (DenseNet) as the first part of such a model and use highway connectivity in the DNN component. This allows us to build a deeper architecture that can learn discriminative emotional contexts in a robust way.  We also investigate the effects of different data augmentation techniques -- especially, the recently proposed mixup technique on the robustness of systems. The proposed model is evaluated and compared with popular architectures in diverse situations including noisy environment, adversarial attacks, and cross-corpus SER. We found that the proposed model shows improved robustness compared to the state-of-the-art models.  
% %BS: You could say on which databases

\end{abstract}

\section{Introduction}
Despite the significant progress in Speech Emotion Recognition (SER) through Deep Neural Networks (DNNs), SER systems still %BS: 
perform poorly in noisy environments \cite{pandharipande2019front,triantafyllopoulos2019towards}, and when the imperceptible adversarial perturbation is added to test examples \cite{latif2018adversarial}. The performance of state-of-the-art SER also degrades in the cross-corpus setting when an acoustic mismatch between training and testing exists. This shows that SER systems lack robustness and generalisation which makes them susceptible to unknown test data shifts. Researchers have developed various methods to improve the performance of SER in noisy environment \cite{triantafyllopoulos2019towards,latif2020deep} and cross-corpus setting \cite{latif2018transfer}, however, significant performance improvement is still required. %Sensing technology \cite{rana2016gait} 

Robustness in DNNs can be enhanced by utilising very deep architectures. Many layers in a deep architecture allow learning very complex patterns, therefore anomalies through the noise and other conditions imposed by cross-corpus can be isolated relatively easily. Studies on speech have shown that utilisation of very deep architectures has led to robustness against noisy situations by learning complex representations \cite{qian2016very,tan2018adaptive}. However, deep networks are difficult to train as it is subject to falling into local extrema, also takes longer time and powerful computational resources, e.g. GPU. Very recently, deep networks like DenseNet \cite{huang2017densely} and highway networks \cite{srivastava2015highway} have gained popularity,  as they allow for easy training of very deep feedforward networks with considerably fewer parameters. DenseNets are densely connected Convolutional Networks and Highway networks closely follow the structure of the Long Short-Term Memory through gating mechanisms. 
% These networks learn deep representations that help achieve better performance and turn out to be more robust \cite{zhu2018sparsely}
Several studies in vision and ASR have shown the benefit of using DenseNet and highway networks, however, their performances for SER need to be evaluated. SER introduces the temporal dimension differentiating it from images. It is also complicated than ASR, as ASR mainly deals with verbal messages, but SER works on vocal expressions that mesh verbal messages coded in an arbitrary and categorical fashion with nonverbal affect signalling system coded in an iconic and continuous fashion~\cite{juslin2008speech}.

Robustness can also be enhanced using data augmentation techniques, which improve the generalisation and help DNNs to provide robustness against unseen real-time situations \cite{hendrycks2019augmix,zhang2017mixup}. Recently ``mixup''~\cite{zhang2017mixup} shows great promise for data augmentation by augmenting a synthetic sample as a linear combination of the original sample. It can make the training data more diverse and the regularisation effect more powerful, so as to further improve the generalisation ability of the network~\cite{liang2018understanding}. Despite its potential, the performance of mixup is not validated for SER.

Besides noise, several studies have shown that SER systems are also susceptible to adversarial attacks and their performance can significantly drop due to the attack \cite{latif2018adversarial,gong2017crafting}.  Adversarial attacks  are developed by malicious adversaries to craft adversarial examples by the addition of unperceived perturbation to elicit wrong responses from machine learning (ML)
models. Methods to achieve robustness against the adversarial attacks in SER systems have not been evaluated.

This paper makes several contributions.
\begin{enumerate}
\item  We propose a deep SER model built on DenseNet and highway networks for robust SER.
\item To further improve robustness, we propose mixup as a data augmentation method.
\item We comprehensively evaluate the robustness of the proposed model in three distinct settings (a) noisy conditions, (b) adversarial attacks, and (c) cross-corpus scenarios.
\end{enumerate}

\section{Related Work}
In recent years, several studies in the image domain \cite{guo2019meets} and in Automatic Speech Recognition (ASR) have used deep architectures to achieve robustness. In \cite{tan2018adaptive}, the authors 
used a very deep convolutional residual network (VDCRN) for noise-robust speech recognition. They empirically showed VDCRN is more robust to noise and able to significantly reduce the word error rate (WER). Studies in vision \cite{zhu2018sparsely,guo2019meets} showed DenseNet is more robust compared to the other state-of-the-art models including ResNet \cite{he2016deep} due to its efficient feature reuse ability.  Similarly, studies \cite{strake2018densenet,9054499} have found that DenseNet can help achieve robustness and generalisability in ASR. However, none of these studies evaluates DenseNet for robust SER.  In this work, we modify DenseNet architecture and use it as a feature extractor in our proposed model.

Although we could not find any study using very deep architectures for SER, several studies have considered DNNs to achieve robustness for SER. Huang et al.\  \cite{huang2017deep} used a convolutional neural network (CNN) -- long-short term memory (LSTM) 
CNN-LSTM model for robust SER. They found that CNN demonstrates a certain degree of noise robustness. However, this study does not utilise very deep architectures. In \cite{triantafyllopoulos2019towards}, the authors utilised deep residual networks as an enhancement architecture to remove noise from speech while preserving enough information for an SER system. This study was focused on speech enhancement instead of robust representation learning. Some studies \cite{pandharipande2019front,juszkiewicz2014improving} also explored different noise removal methods for SER in noisy environments.

% , however, no study evaluated deep architectures to achieve robustness in SER. 

%\RR{The following paragraph is more appropriate for an introduction. You should bring papers that used data augmentation for avoiding adversarial attacks. Even from other domains is ok, if you do not have any paper in SER}
Data augmentation is a well-known practice to enlarging the size of the training set in many machine-learning applications.  Recently, it has been shown that mixup data augmentation can enhance the classifiers' robustness for unseen test data \cite{zhang2017mixup}. It also improves generalisation performance and model robustness against adversarial examples \cite{pang2019mixup}. The regularisation effect of mixup has been evaluated for ASR \cite{tomashenko2018speaker}, however, no study has evaluated mixup in SER. Here, we use mixup to improve generalisation and robustness of proposed model against noisy environment, adversarial attacks, and cross-corpus setting.

%Studies \cite{zantedeschi2017efficient} have also used data augmentation techniques to improve the robustness against adversarial attacks \cite{zantedeschi2017efficient}. 
%Therefore, in this work, we evaluate data augmentation techniques for SER in noisy environment, adversarial attacks, and cross-corpus setting.  

%Therefore, in this work, we aim to build a robust model that can perform SER in more challenging situations including noisy environment, against adversarial attacks, and cross-corpus setting. 

% % For this, we design a new SER model by utilising 
% %BS: 
% DenseNet 
% and highway 
% %BS: 
% architectures 
% to learn robust representations. We also use different data augmentation techniques to show their effectiveness towards system robustness. 
%BS: Some here feels repetitive - on the other hand, mixup appears for the first time. I also wonder whether it could be good to be more specific in the title by either including ": On noise, cross-corpus settings, and adversarial attacks" or by including ": Dense and Highway Nets and Mixup Data Augmentation".
%BS: If more space is needed, the intro/related work part can be shortened removing redundancy.

% However, methods to achieve robustness against these attacks in SER systems has not been evaluated.  Therefore, in this work we also use different data augmentation techniques to show their effectiveness towards system robustness. Particularly, we use recently proposed mixup \cite{zhang2017mixup} data augmentation technique that has shown promising results in achieving robustness.  

\section{Proposed Model}
\begin{figure}[!t]
  \centering
  \includegraphics[trim=0cm 0cm 0cm 0cm,clip=true,,width=0.43\linewidth]{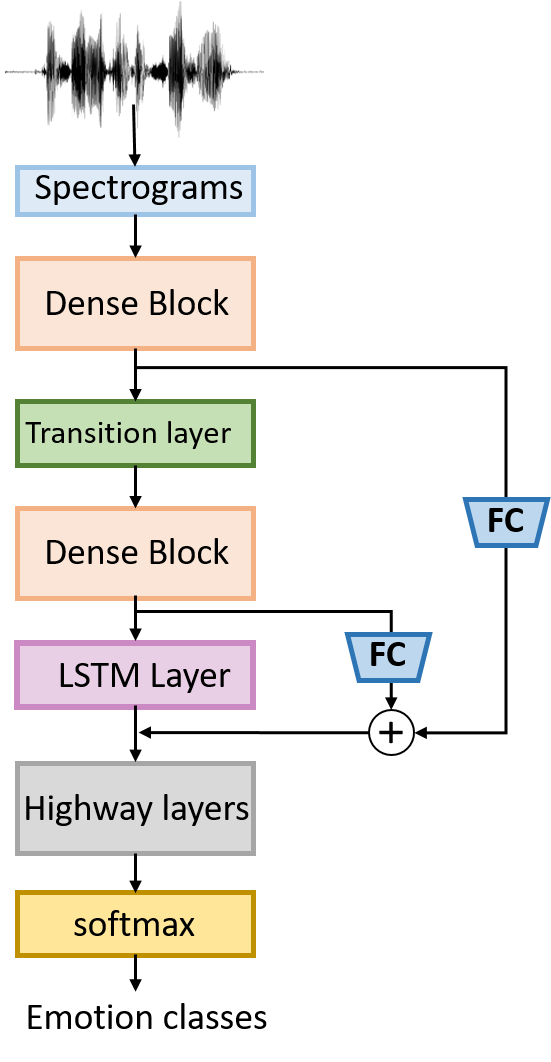}
  \caption{Schematic diagram of the proposed model.}
  \label{fig:model}
\end{figure}

% \RR{Comment:This section should describe the model first before describing its functions.\newline}
Our proposed model is a hybrid architecture, where we use a DenseBlock for temporal feature extraction, LSTM for context aggregation and fully connected layers in highway configuration for discriminative feature learning. A schematic diagram of the proposed model is shown in Figure~\ref{fig:model}. 

% In this section we describe how the depth of the architecture enables robustness.

% To improve robustness, we propose a model that can learn deep emotional representations. It consists of different building blocks that help each other in achieving robust SER. The details of these building blocks are given below. 
% trim={<left> <lower> <right> <upper>}
% trim=2.82cm 0.3cm 1.2cm 0.3cm,clip=true,
%BS: PLEASE DO NOT embed images as bitmaps such as png - please use scalable pdf or postscript instead - thanks :)

\subsection{Temporal Feature Capturing using DenseNet}
%BS: 
The first element of our model is DenseNet. DenseNet enables learning temporal features 
by introducing direct connections from each layer to all subsequent layers. Consequently, the $l^{th}$ layer ($x_l$) receives the feature maps of all preceding layers as input: 
\begin{equation}
\label{con}
    x_l=H_{l,G}([x_o, x_1,...,x_{l-1}])
\end{equation}
Here, $H_{l,G}(.)$ represents to a composite function including batch normalisation (BN) \cite{ioffe2015batch}, a rectified linear unit (ReLU) \cite{nair2010rectified} and a convolution (Conv) layer. $G$ is the growth rate that represents the number of output feature maps. Cascading multiple layers of composite functions and feature map concatenations forms a so-called DenseBlock $(L, G)$, which has $L$ layers and a growth rate of $G$. The concatenations (Equation \ref{con}) in the DenseBlock causes the input size to be increased as the number of layers increases in DenseBlock. For downsampling, a transition layer is used after each DenseBlock. 
The transition layer consists of a batch normalisation layer, a $1\times1$ convolutional layer and a
$2\times2$ average pooling layer. 

In the DenseNet architecture, there is a global average pooling layer that we replace with a reshape layer to adapt the dimension of feature maps to LSTM layer for context learning. % However, in our baseline model, we use the global average before the classification.} 

\subsection{Emotional-context Modelling using LSTM}
Emotions are context-dependent \cite{Latif2018} and contexts are embedded in temporal dimension of the data \cite{latif2019direct}.
LSTM \cite{hochreiter1997long} have gated architectures, which enabling memory, helps to model temporal relationships. We, therefore, use LSTM after the DenseBlock for context modelling.

\subsection{Discriminative Feature Extraction using Highway Network}
Finally, we use the highway network \cite{srivastava2015highway} to extract high-level discriminative features. We also use two skip 
connections from each DenseBlock and the sum of these features are concatenated with the LSTM features and given to the highway network. Skip connections introduce the shortcut connection that shuttles different levels of abstractions and also improves the gradient flow 
in the network \cite{tong2017image}. The output $y$ of a highway block is given by:
\begin{equation}
\label{highww}
y= H(x, W_{H}).T(x,  W_{T})+x.C(x W_{C}),    
\end{equation}
where $H$ (parameters by $W_{H}$) is a nonlinear transformation on its associated input $x$, and $C$ and $T$ represent carry gates and transform gates, respectively. The layer indices and biases have been excluded in Equation \ref{highww} for simplification. 

Gates in a highway network control information flow and speed up convergence enabling effective training of DNNs across several layers without degradation \cite{srivastava2015highway}. This helps the highway network to learn high-level discriminative representation \cite{wang2017tacotron} that help to achieve better classification performance. Therefore, we use the highway network layers prior to softmax layer for classification.

 \section{Experimental Setup}
\subsection{Dataset}
We evaluate our model on two popular datasets: Interactive Emotional Dyadic Motion Capture (IEMOCAP) \cite{busso2008iemocap} and MSP-IMPROV \cite{busso2017msp}. The detail of these datasets is given below. \linebreak
\textbf{IEMOCAP:}
This corpus contains five sessions, where each session has utterances from two speakers (one male and one female). Overall, there are 10 unique speakers. We consider  four emotions including angry, happy, neutral and sad. To be consistent with previous studies \cite{aldeneh2017using}, we merge excitement with happiness and consider it as one class: happy.\linebreak
\textbf{MSP-IMPROV:}
The MSP-IMPROV dataset contains six sessions, where each session comprises of utterances from two speakers, one male, and one female. There are four emotion categories in MSP-IMPROV: angry, neutral, sad, and happy. All were used in the experiments.\linebreak
\textbf{DEMAND}
We choose the Diverse Environments Multichannel Acoustic Noise Database (DEMAND) dataset \cite{thiemann2013diverse} as a source of our noise signal. This data contains the recording of various real-world noises in a variety of settings. We select 
noise recordings of 16\,kHz sampling rates.

\subsection{Data Pre-processing}
We use spectrogram as our starting point. We compute them using Short-Time Fourier Transform (STFT) with an overlapping Hamming window of size 25\,ms with a 10\,ms shift. We select the height of spectrogram equal to 128. We apply a context window of 256 frames to reach a 
fixed width of segments of spectrograms following the procedure used in \cite{latif2020multi}. Each segment is assigned the emotion labelling of the corresponding utterance. We train all the models using short segments, however, utterance level prediction is calculated by averaging the posterior probabilities of the respective sub-segments.

\subsection{Data Augmentation}
We use the ``mixup'' data augmentation technique, which has not been used in SER. It creates training samples using following equations: 
\begin{align}
& \tilde{x}=\lambda x_{i}+(1-\lambda) x_{j} \label{m1} \\
&\tilde{y}=\lambda y_{i}+(1-\lambda) y_{j}\label{m2}
%BS:
.
\end{align}
Here, ($x_{i}$, $y_{i}$) and ($x_{j}$, $y_{j}$) are two randomly selected examples from training data, and $\lambda$ $\in$ [0, 1]. Mixup can be employed on raw speech as well as on features \cite{zhang2017mixup}. We use mixup on spectrograms.

We also use the widely used speed perturbation (SP) for data augmentation. We follow \cite{latif2019direct} to create samples using speed perturbation.  For a given utterance, we produce two versions of each utterance by applying the speed effect at the factors of 0.9 and 1.1. 

 \subsection{Model Configuration}
 \label{Model_C}
In our DenseNet architecture, after the initial convolutional layer, we use two DenseBlocks and each DenseBlock consists of $L$ = 6 layers with a growth rate of $G$ = 24. After the first DenseBlocks, we place a transition layer for down-sampling the size of feature maps. This consist of a 1$\times$1 convolutional layer followed by a 2$\times$2 average pooling layer with stride of 2. The transition of temporal features from the DenseBlock to LSTM is made using a reshape layer. The LSTM layer serves to learn contextual information from the temporal features extracted by the DenseBlocks. After the LSTM layer, we insert a  dropout layer with a dropout value of 0.5. The sum of features discovered by each DenseBlock, having two skip connections, is concatenated with the contextual features from LSTM and given to the highway network to learn discriminative features. We apply three fully connected layers with 128 hidden units with ReLU activation in the highway block followed by a softmax layer for emotion classification.  

We also implement benchmarking models including DenseNet, DenseNet-LSTM, CNN, and CNN-LSTM. In DenseNet, we apply three DenseBlocks and each consists of $L$ = 6 layers with a growth rate of $G$ = 16. After the DenseBlocks, we place a 3$\times$3 global average pooling layer and a fully connected layer of 1\,000  hidden units before a softmax layer. 

For DenseNet-LSTM, we use an LSTM layer instead of a global average pooling layer. In all convolutional layers used in the DenseNet based models, batch normalisation \cite{ioffe2015batch} is employed before the non-linearities. A rectified linear unit (ReLU) \cite{nair2010rectified} is used as the activation function. 

For the CNN-LSTM, we follow the architecture configuration described in \cite{satt2017efficient}. To use CNN for benchmarking, we choose a fully connected layer instead of an LSTM layer in the above CNN-LSTM model.

We train the models using the training set and validation set is used for hyper-parameter selection. For minimisation of the cross-entropy loss function, we choose the Adam optimiser \cite{kingma2014adam}. We start the training with an initial learning rate of $10^{-3}$. If the validation accuracy does not improve for five consecutive epochs, we halve the learning rate and stop the process if the validation accuracy does not improve for 20 consecutive epochs. For each model, we repeat the evaluation 10 times and report the mean and standard deviation.

\section{Experiments and Results}
We apply a leave-one-speaker-out scheme for both datasets and report unweighted average recall (UAR) for both datasets. UAR is a widely used metric used for speech emotion recognition due to the dominance of class imbalanced datasets in this field. In each session, we employ utterances from one speaker for testing and the other speakers' utterances for validation.  This configuration is used for all models. Results are computed in noisy, cross-corpus, and adversarial attacks settings. 

\subsection{Benchmark Results}
The comparison of our proposed model with the benchmark models CNN, CNN-LSTM, DenseNet and DenseNet-LSTM are presented in Table \ref{bech}. All these results are computed without data augmentation. We observe that for both IEMOCAP and MSP-IMPROV datasets, our proposed model performs better than the benchmarking models.  
\begin{table}[!ht]
\centering
\scriptsize
\caption{UAR (\%) of different models. }
\begin{tabular}{l|l|l}
\hline
Model       & IEMOCAP & MSP-IMPROV \\ \hline
CNN         &  61.5 $\pm$ 2.3     & 52.6 $\pm$  2.5        \\ \hline
CNN-LSTM    &  62.1 $\pm$1.8      & 53.1 $\pm$ 2.3         \\ \hline
DenseNet    &  63.2 $\pm$ 1.7     & 54.5 $\pm$ 1.9        \\ \hline
DenseNet-LSTM & 63.5  $\pm$ 1.5    &  55.6 $\pm$  1.6      \\ \hline
Proposed    &  \textbf{64.1 $\pm$ 1.3}    &  \textbf{56.2  $\pm$ 1.5}     \\\Xhline{1pt}
CNN-LSTM \cite{satt2017efficient}   &  62.0     &  --     \\ \hline
CNN \cite{latif2020multi}   &  61.4    &  55.3    \\\Xhline{1pt}
\end{tabular}
\label{bech}
\end{table}

Our proposed model is also performing better compared to standard DenseNet and DenseNet-LSTM, which shows that the use of highway connectivity in our proposed model is helping to achieve better results compared to the variants of CNNs, DenseNet, and recent studies \cite{satt2017efficient,latif2020multi}.
\subsection{Noisy Environment}
To evaluate the model in a noisy environment, we select three signal-to-noise ratio (SNR) values [0, 10, 20]. We consider the mismatched condition, where the model is trained on clean data and the test data incorporates noisy samples. We choose five noises including kitchen, park, station, traffic, and cafeteria from the DEMAND dataset. These noises are randomly added to the test set at three different SNR levels. Results on IEMOCAP data are reported in Table \ref{noise}. It can be noted from Table \ref{noise} that the proposed model provides better results compared to all the other models.
\begin{table}[!t]
\centering
\scriptsize
\caption{UAR (\%) of different models on IEMOCAP data in noisy environment setting.}
\begin{tabular}{c|c|c|c|c}
\hline
\multirow{2}{*}{\diagbox{\tiny{SNR (dB)}}{\tiny{Model}}} & \multirow{2}{*}{\tiny{CNN-LSTM}} & \multirow{2}{*}{\tiny{DenseNet}} & \multirow{2}{*}{\tiny{DenseNet-LSTM}} & \multirow{2}{*}{\tiny{Proposed}} \\&                           &                          &                               &                           \\ \hline
0       &  21.0$\pm$ 2.5    &     22.3$\pm$ 1.8    &   23.1$\pm$  1.6      &  \textbf{24.4$\pm$  1.5}    \\\hline
10    &    28.2 $\pm$ 2.1   &  30.0 $\pm$ 2.0          &  31.4 $\pm$ 1.8      &  \textbf{32.3 $\pm$ 1.1}           \\ \hline
20       & 34.8  $\pm$ 1.8    &  35.5  $\pm$ 1.4        &    36.6 $\pm$ 1.6   &  \textbf{37.1 $\pm$ 1.6}                         \\ \hline
\multicolumn{5}{c}{speed perturbation}                                                                                                                          \\ \hline
0  &   26.5$\pm$ 2.0        &  28.5 $\pm$ 1.5    &  30.2 $\pm$ 1.3  &  \textbf{31.7 $\pm$ 1.4}                         \\ \hline
10  &  32.5$\pm$ 1.9    &    32.9$\pm$ 1.3      &33.1$\pm$ 1.4&  \textbf{34.5$\pm$ 1.5}                         \\ \hline
20 &   38.9  $\pm$ 1.4      &    39.1  $\pm$ 1.6     &  40.2  $\pm$ 1.2       & \textbf{40.8 $\pm$ 1.0}                           \\ \hline
\multicolumn{5}{c}{mixup}                                                                                                                                       \\ \hline
0  &  28.1$\pm$ 2.4  &    30.8 $\pm$ 1.2   & 31.1 $\pm$ 1.1       &   \textbf{32.5$\pm$ 1.1}                       \\ \hline
10 &   33.6$\pm$ 1.8   & 34.7$\pm$ 1.5      &  34.5$\pm$ 1.4          & \textbf{34.9$\pm$ 1.6}                          \\ \hline
20 &   39.2$\pm$ 1.4  & 40.8  $\pm$ 1.3       &  41.6  $\pm$ 1.3    &    \textbf{42.5  $\pm$ 1.3}                         \\ \hline

\multicolumn{5}{c}{speed perturbation+mixup}                                                                                                                                       \\ \hline
0  &  30.3$\pm$ 2.0  &    33.9 $\pm$ 1.4   & 33.7 $\pm$ 1.0       &   \textbf{34.2$\pm$ 1.2}                       \\ \hline
10 &   35.2$\pm$ 1.2   & 38.8$\pm$ 1.2      &  40.6$\pm$ 1.2          &\textbf{ 40.9$\pm$ 1.5}                          \\ \hline
20 &   40.6$\pm$ 1.3  & 42.0  $\pm$ 1.1       &  41.8  $\pm$ 1.5    &   \textbf{ 43.1 $\pm$ 1.1 }                        \\ \hline

\end{tabular}
\label{noise}
\end{table}

We also observe from Table \ref{noise} that the data augmentation techniques help to improve robustness. Data augmentation using mixup achieves better results compared to that using speed perturbation, however, the combination of mixup and speed perturbation provides the best results.  

\subsection{Adversarial Settings}
In adversarial settings, we use two adversarial attacks including the Fast Gradient Sign Method (FGSM) \cite{goodfellow6572explaining} and the Basic Iterative Method (BIM) \cite{kurakin2016adversarial} to evaluate the robustness. FGSM creates the adversarial examples by adding a scaled noise in the direction of the gradient of the loss function.  Instead of applying adversarial noise in a single step like FGSM, BIM iteratively applies it multiple times. We applied these two attacks with the perturbation factor $\epsilon=0.08$ on different classifiers and performance is reported in Table \ref{Adv}. We observe that our proposed model performs the best with or without data augmentation. We also observe data augmentation techniques help to improve robustness in the adversarial settings in the same way as in the presence of noise.
\begin{table}[!t]
\centering
\scriptsize
\caption{UAR (\%) of different models on IEMOCAP data in adversarial setting. }
\begin{tabular}{c|c|c|c|c}
\hline
\multirow{2}{*}{\diagbox{\tiny{Attack}}{\tiny{Model}}} & \multirow{2}{*}{\tiny{CNN-LSTM}} & \multirow{2}{*}{\tiny{DenseNet}} & \multirow{2}{*}{\tiny{DenseNet-LSTM}} & \multirow{2}{*}{\tiny{Proposed}} \\&                           &                          &                               &                           \\ \hline
FSGM       &  30.1$\pm$ 1.8    &     33.5$\pm$ 1.4    &   34.5$\pm$  1.7      &  \textbf{35.2$\pm$  1.3}    \\\hline
BIM    &    27.2 $\pm$ 2.1   &  29.7 $\pm$ 1.8          &  31.4 $\pm$ 1.4      &  \textbf{32.8 $\pm$ 1.1}           \\ \hline
\multicolumn{5}{c}{speed perturbation}                                                                                                                          \\ \hline
FSGM  &   34.5$\pm$ 2.0        &  38.2 $\pm$ 1.4    &  38.1 $\pm$ 1.2  &  \textbf{40.2 $\pm$ 1.0}                        \\ \hline
BIM  &  30.5$\pm$ 1.9    &    33.5$\pm$ 1.7     &33.9$\pm$ 1.4&  \textbf{34.6$\pm$ 1.2}                         \\ \hline
\multicolumn{5}{c}{mixup}                                                                                                                                       \\ \hline
FSGM  &  36.1$\pm$ 2.4  &    39.8 $\pm$ 1.7   & 40.5 $\pm$ 1.2      &   \textbf{41.4$\pm$ 1.4}                       \\ \hline
BIM &   31.3$\pm$ 1.8   & 34.4$\pm$ 1.2     &  34.0$\pm$ 1.4          &\textbf{34.8$\pm$ 1.2}                          \\ \hline

\multicolumn{5}{c}{speed perturbation+mixup}                                                                                                                                       \\ \hline
FSGM  &  39.1$\pm$ 2.4  &    42.8 $\pm$ 1.6   & 42.5 $\pm$ 1.6       &   \textbf{44.0$\pm$ 1.1}                       \\ \hline
BIM &   32.6$\pm$ 1.8   & 35.4$\pm$ 1.2     &  36.8$\pm$ 1.2          & \textbf{37.4$\pm$ 1.3}                          \\ \hline
\end{tabular}
\label{Adv}
\end{table}
\subsection{Cross-Corpus Settings}
\begin{figure}[!t]
  \centering
  \includegraphics[trim=0cm 0.5cm 0.8cm 0.1cm,clip=true,width=\linewidth]{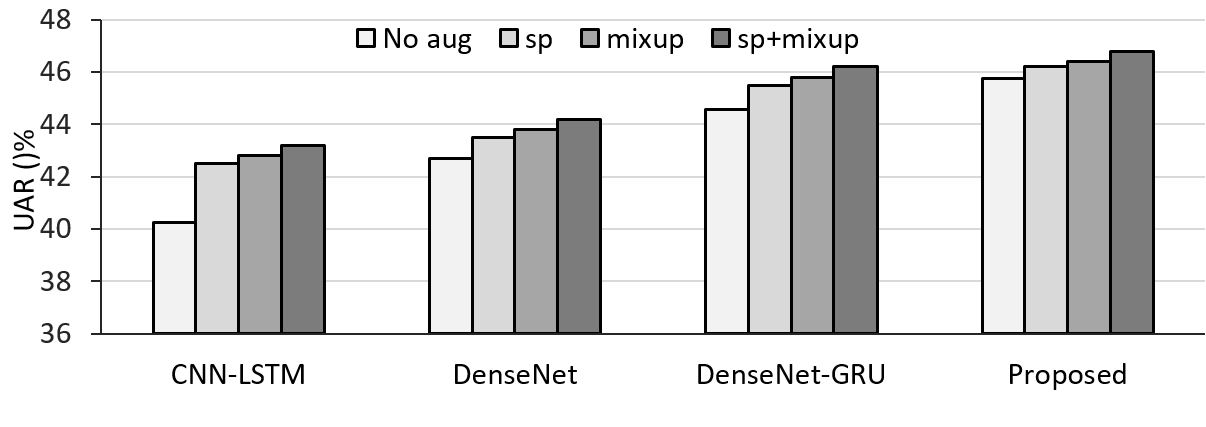}
  \caption{Comparing different models in cross-corpus SER.}
  \label{fig:crosscorpus}
\end{figure}

To evaluate the proposed model in a cross-corpus setting, we use IEMOCAP as training data and MSP-IMPROV as the test set. 
We randomly select 30\,\% of MSP-IMPROV for parameter selection and 70\,\% for testing, as used in \cite{latif2020multi}. We evaluate different models in the cross-corpus setting. The results are given in Figure \ref{fig:crosscorpus}. We observe that the proposed model achieves better performance and data augmentation helps to improve the robustness. 
% trim={<left> <lower> <right> <upper>}
\begin{table}[!t]
\centering
\scriptsize
\caption{Comparing cross-corpus results with recent studies.}
\begin{tabular}{l|l|l|l}
\hline
Studies  &Latif et al.\ \cite{latif2020multi}  & Sahu et al.\ \cite{sahu2019multi}  & Proposed \\ \hline
UAR (\%) &46.41$\pm$0.32  & 40.08 & \textbf{46.81$\pm$0.40} \\ \hline
\end{tabular}
\label{cross-comp}
\end{table}

We also compare our results with previous studies (\cite{latif2020multi,sahu2019multi},) in the cross-corpus setting in Table \ref{cross-comp}. In~\cite{latif2020multi}, authors employ a multi-task framework and exploit larger unlabelled data for the auxiliary task to improve the generalisation of the model. In \cite{sahu2019multi}, the authors develop a multi-modal technique (audio plus text) for SER based on ASR transcriptions. They demonstrate that the generalisability of ASR models helps to improve the generalisation of emotion classification models. In contrast, we propose a deeper model coupled with data augmentation to achieve improved generalisation. We are achieving better results compared to these studies as reported in Table \ref{cross-comp}.

\section{Conclusions}
This paper introduces a new hybrid model to build a robust SER system. This model exploits a DenseNet for feature extraction, LSTM for contextual learning, and deep neural network layers with highway connectivity for discriminative representation learning, and produce robust representation. This paper also proposes data augmentation to further improve the robustness of the architecture. The performance of our proposed technique is evaluated on widely used IEMOCAP and MSP-IMPROV datasets against noise, adversarial attacks, and cross-corpus settings. Results show that our proposed technique is more robust compared to existing methods and other state-of-the-art models. Results also reveal several valuable information, such as, mixup is a better augmentation technique for SER compared to the popular speed perturbation. Results also show that DenseNet based models are more robust compared to CNN-LSTM or just CNN. 
In future work, we aim at further optimising these architectures and augmentation in closer loop.

% \bibliographystyle{IEEEtran}

% \bibliography{mybib}

% Generated by IEEEtran.bst, version: 1.13 (2008/09/30)

\end{document}